\documentclass[reprint,twocolumn,showpacs,preprintnumbers,amsmath, aps,pre,amssymb]{revtex4-1}
\usepackage{algorithmic}
\usepackage{algorithm}
\usepackage{ amsmath}
\usepackage{graphicx}% Include figure files
\usepackage{dcolumn}% Align table columns on decimal point
\usepackage{bm}% bold math

\newcommand{\remove}[1]{} % comment out text

\begin{document}

\preprint{APS/123-QED}

% Dynamics of Synchronization in a Network of Oscillators Coupled vis Non-Conservative Interactions
% A Generalized Model of Synchronization in Complex Networks
% Dynamics and Topological Scales of a Generalized Model of Synchronization in Complex Networks
%\title{Dynamics of Generalized Models of Synchronization in Complex Networks}
%\title{Role of Dynamics in Network Structure in Generalized Models of Synchronization}
\title{Network Structure, Topology and Dynamics in Generalized Models of Synchronization}

\author{Kristina Lerman}
% \email{lerman@isi.edu}
\author{Rumi Ghosh}
% \email{rumig@usc.edu}
\affiliation{%
USC Information Sciences Institute\\
4676 Admiralty Way, Marina del Rey, CA 90292
}%

\date{\today}% It is always \today, today,

\begin{abstract}
%Network structure is a product of the connectivity of its nodes and %the dynamic processes taking place on it.
%interactions between them.
We explore the interplay of network structure, topology, and dynamic interactions between nodes using the paradigm of distributed synchronization in a network of coupled oscillators. As the network evolves to a global steady state, interconnected oscillators synchronize in stages, revealing network's underlying community structure. Traditional models of synchronization assume that interactions between nodes are mediated by a conservative process, such as diffusion. However, social and biological processes are often non-conservative. We propose a new model of synchronization in a network of oscillators coupled via non-conservative processes.
We study dynamics of synchronization of a synthetic and real-world networks
and show that different synchronization models reveal different structures within the same network.
\end{abstract}

\pacs{05.45.Xt, 89.75.Hc, 89.75.-k, 89.65.Ef, 89.75.Fb, 02.10.Ud}% PACS, the Physics and Astronomy
\maketitle

{Community structure} is an important characteristic of complex networks, including biological and social networks which are composed of functional modules or groups  of similar individuals~\cite{Ravasz2002Hierarchical,Rives03}. Existing algorithms examine connectivity, or topology, of the network to partition it into communities~\cite{Chung1997Spectral,Spielman96spectralpartitioning,Fortunato2010Community}. However, community structure of real-world networks is the product of both their topology \emph{and function}, which is determined by dynamic processes taking place on the network. These processes are mediated by interactions between nodes, and they determine the phenomena taking place on the network, whether diffusion and other types of transport in biological networks, or epidemics and information spread in social networks. While it is generally accepted that network structure affects the evolution of dynamic phenomena~\cite{Strogatz01,Boccaletti06}, the impact of dynamics  on our understanding of structure is less appreciated.
As we show in this paper, different dynamic processes running on the same network can lead to different views of network structure.

We explore the connection between network structure, topology and dynamic processes by studying synchronization in a network of coupled oscillators.
%We explore the connection between network topology, dynamic processes and structure using synchronization paradigm.
Kuramoto~\cite{Kuramoto} introduced a simple model of distributed synchronization that was adapted to networks of oscillators whose phases are coupled to their neighbors' phases~\cite{Nishikawa03,Boccaletti06}.
These systems demonstrate an interesting connection between dynamics and structure: as the network evolves to a steady state, oscillators belonging to different communities synchronize in stages, revealing the network's hierarchical community structure~\cite{Arenas06,Arenas2008Synchronization}.

%In the Kuramoto model, oscillators are coupled via a conservative process similar to diffusion.
Oscillators in the Kuramoto model are coupled via a processes similar to diffusion, which we informally refer to as conservative.
Such processes cannot describe interactions in real-world networks, which often have non-conservative nature, for example, due to dissipation.
%Real-world networks, however, are often non-conservative. As an epidemic spreads through a social network, the amount of the pathogen will increase, and as individuals are cured, decrease. Biological processes are also prone to dissipation.
To account for this, we introduce a new model of synchronization in a network of nodes coupled via non-conservative processes.
We simulate the two types of processes in example networks: a synthetic network with a hierarchical community structure and a benchmark social network. We show that dynamics of non-conservative synchronization  reveals a community structure, but this structure is different from that found by the conservative Kuramoto model. This demonstrates the importance of dynamic processes in the analysis of network structure.

We consider a network of $N$ active nodes, each interacting locally with its neighbors. In the Kuramoto model nodes are oscillators coupled to their neighbors through the sine of their phase differences. The phase $\theta_i$ of the $i$th oscillator evolves in time according to:
\begin{equation}
\frac{d\theta_i}{dt} =\omega_i+\sum_{j \in neighb( i)} K_{ij}  sin(\theta_j-\theta_i)  \label{eq:31}
\end{equation}
\noindent where $\omega_i$ is the natural frequency of node $i$, and $K_{ij}$ is the coupling constant that describes the strength of interaction with neighbor $j$. For small phase differences,  $sin \theta \approx \theta$, the linear version of the Kuramoto model can be written in vector form:
\begin{equation}
\label{eq:1}
\frac{d\theta}{dt}=\omega - K\cdot L\theta
\end{equation}
\noindent Here $\theta$ is a vector of phases and $\omega$ natural frequencies of $N$ oscillators, and $K$ is a matrix of pairwise couplings constants between nodes. More generally, $\theta$ can be some dynamic variable associated with the nodes.  We represent the network as an unweighted, undirected graph, with an adjacency matrix $A$, such that $A[i,j]=1$ if there exists an edge between nodes $i$ and $j$; otherwise, $A[i,j]=0$. The Laplacian matrix of the graph is $L=D-A$, where $D$ the diagonal degree matrix, such that $D[i,i]=\sum_i A[i,j]=d_i$ and $D[i,j]=0$ $\forall$ $i\neq j$.
If nodes are identical, $\omega_i=\omega$ $\forall i$, the network can reach a fully synchronized steady state in which  $\theta_i=\theta$ $\forall i$.

A simple intuition for why the interactions in the linear Kuramoto model may be considered conservative is as follows. Imagine that at time $t$, node $i$ produces an amount $d_i\theta_{i}(t)$ of some quantity (e.g., chemical density) for its $d_i$ neighbors. Each neighbor receives $1/{d_i}$ of this amount; therefore, interaction conserves the amount of quantity in the network.
Interactions need  not always be conservative. In some networks, the quantity created by a node (e.g., pathogen in a human population, light among fireflies)  is not completely distributed among neighbors. Imagine now that node $i$ produces an amount $\alpha \theta_i$ of the quantity for its neighbors, regardless of the number of neighbors it has, but each neighbor receives an amount $\theta_{i}$. Therefore, $(\alpha-d_i)\theta_i$ of the quantity created by $i$ is not transferred to any neighbor and is lost.
This changes the nature of interactions and the resulting network dynamics.
A linear model of non-conservative interactions described above can be written as:
\begin{equation}
\frac{d\theta_i}{dt} =\omega_i+\sum_{j \in neighb( i)} K_{ij}  \big(\theta_j-\frac{\alpha \theta_i}{d_i}\big),\label{eq:4}
\end{equation}
\noindent Here $\alpha \ge \lambda_{max}$ is a constant,  with $\lambda_{max}$ the largest eigenvalue of $A$.
%In contrast, in the linear Kuramoto model, node $i$ produces an amount $d_i\theta_{i}(t)$ of the quantity for its neighbors. Each neighbor then receives $1/{d_i}$ of this amount; therefore, interaction conserves the amount of quantity in the network.
Equation~\ref{eq:4} can be written in vector form: ${d\theta}/{dt}=\omega -K\cdot(\alpha I-A)\theta$,
where $I$ is the identity matrix. The model above introduces a new operator, the \emph{Replicator} operator $R=\alpha I-A$, a non-conservative counterpart of the Laplacian matrix, which governs the dynamics of the network.
In spite of non-conservation, under  conditions specified below,  the system reaches a steady state where the dynamic variable $\theta$ no longer changes.

% generalized interaction models
Both conservative and non-conservative models are special cases of the generalized linear synchronization model, which can be written in terms of the {operator} $\mathcal{L}(A)$ of the adjacency matrix $A$.
\begin{equation}
 \frac{d\theta}{dt}=\omega-K\cdot \mathcal{L}(A)\theta
 \label{eq:5}
 \end{equation}
Each model encapsulates the details of interactions. For example, if the new amount of content produced by node $i$  is $\theta_i$, instead of  $d_i\theta_i$ in Eq. \ref{eq:1}, the conservative synchronization model then leads to the normalized Laplacian $I-AD^{-1}$ rather than $L$.
Solving Eq.~\ref{eq:5} we get:
\begin{equation}
\theta({t})=(\theta_0-{(K\cdot \mathcal{L})}^{-1}\omega) e^{-K\cdot\mathcal{L}t} +{(K\cdot\mathcal{L})}^{-1}\omega
 \label{eq:8}
 \end{equation}
\noindent with $\theta_0$ the initial phase of the oscillator at $t=0$.
Matrix $\mathcal{L}$ can written as an eigenvalue decomposition $\mathcal{L}= \sum_{i=1}^N{\mathcal{X}[.,i]\lambda_i \mathcal{X}^{-1}[i,.]}$, where $\mathcal{X}$ is the matrix whose $i$th column is the $i$th eigenvector of $\mathcal{L}$ with eigenvalue $\lambda_i$.
%Let $\mathcal{X}$ be an $N\times N$ matrix whose $i$th column $\mathcal{X}[.,i]$ gives the eigenvector of $\mathcal{L}$ corresponding to eigenvalue $\lambda_{i}$. Also, let $\Lambda$ be the diagonal matrix of eigenvalues, $\Lambda[i,i]=\lambda_{i}$, with $\lambda_1 \le \lambda_2\le\cdots \lambda_{max}$ and $\mathcal{Y}={\mathcal{X}}^{-1}$. Therefore $\mathcal{L}= \sum_{i=1}^N{\mathcal{X}[.,i]\lambda_i \mathcal{Y}[i,.]}$.
For $\omega=0$ and $K_{ij}=c$ $\forall i,j$,  Eq.~\ref{eq:8}  can be rewritten as:
 \begin{eqnarray}
\theta({t})&=&\theta_{0}e^{-c\mathcal{L}t} =  \sum_{i=1}^N\mathcal{X}[.,i]e^{-c\lambda_i t}\mathcal{X}^{-1}[i,.]\theta_{0} \nonumber \\
&=&\sum_{i=1}^N\mathcal{X}[.,i]e^{-c\lambda_i t}c_i
 \label{eq:9}
 \end{eqnarray}
\noindent Here $c_i=\mathcal{X}^{-1}[i,.]\theta_{0}$ is a constant.

%\paragraph{Steady State}
A non-trivial steady state $\theta(t \rightarrow \infty) \ne 0$ exists when $\lambda_1=0$.
Under this condition, Eq.~\ref{eq:9} as $t\to \infty$ reduces to $\theta^{s}= \mathcal{X}[.,1]c_1$, with constant $c_1$.
The steady state solutions of different synchronization models can be similarly found.
\begin{description}
\item [$\mathcal{L}=D-A=L$]
In this case $\mathcal{X}[.,1]= \bar{1}$ (vector of 1s). Hence, $\theta_i^{s}=\theta$ $\forall$ $i$, i.e., all oscillators have the same phase in the steady state.
\item [$\mathcal{L}=I-AD^{-1}$] In the steady state, $\theta_i^{s} \propto d[i]$, where $d[i]$ is the degree of node $i$.
\item[$\mathcal{L}=\lambda_{max}I-A=R$]  In the steady state, $\theta^{s}$ $\propto$ eigenvector of the adjacency matrix $A$ corresponding to the largest eigenvalue.
\item[$\mathcal{L}=\alpha I-A $ $\forall \alpha > \lambda_{max}$]  The steady state has a trivial solution $\theta_i^{s} \to 0 \ \forall i$
\end{description}
Hu et al.~\cite{Hu} studied dynamics of a network of nodes coupled via a signalling process, which can be described by an operator $\mathcal{L}(A)=(I+A)$. However, it can be shown that this system will never reach a steady state.

The spectrum of the operator $\mathcal{L}$ gives information about topological and temporal scales of synchronization. In the Kuramoto model, the time to reach steady state is inversely proportional to the smallest positive eigenvalue of the Laplacian, and gaps between consecutive eigenvalues of the $L$ are related to the relative difference in synchronization time scales of different communities~\cite{Arenas06,Arenas2008Synchronization}. While we have not fully explored the spectrum of $R$, it can be shown that time to reach the steady state is inversely proportional to the smallest positive eigenvalue of $R$ (for $\alpha=\lambda_{max}$).

\begin{figure*}[tbh]
\begin{tabular}{@{}c@{}c@{}c@{}c@{}}
\includegraphics[height=0.24\textwidth, width=0.24\textwidth]{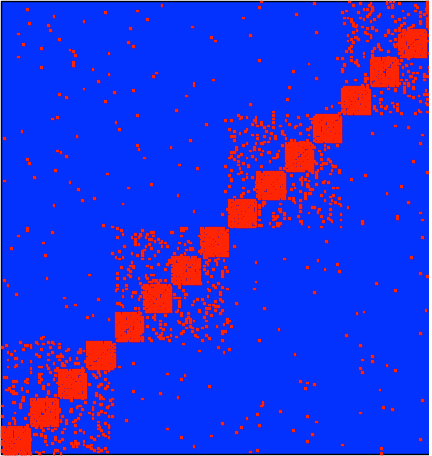} &
\includegraphics[height=0.26\textwidth, width=0.26\textwidth]{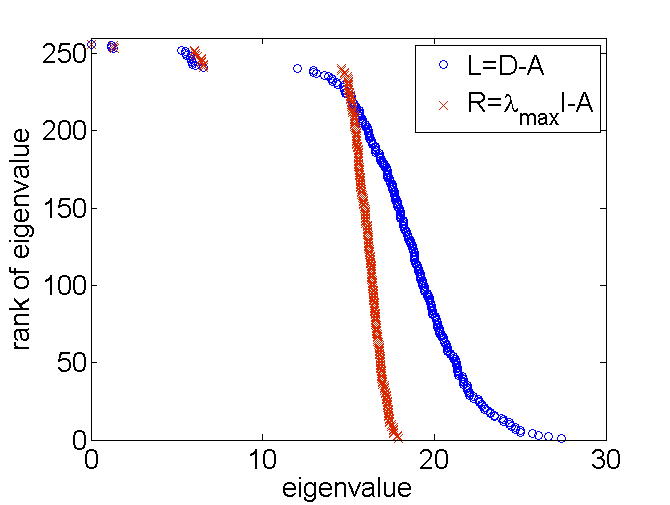} &
\includegraphics[height=0.25\textwidth, width=0.25\textwidth]{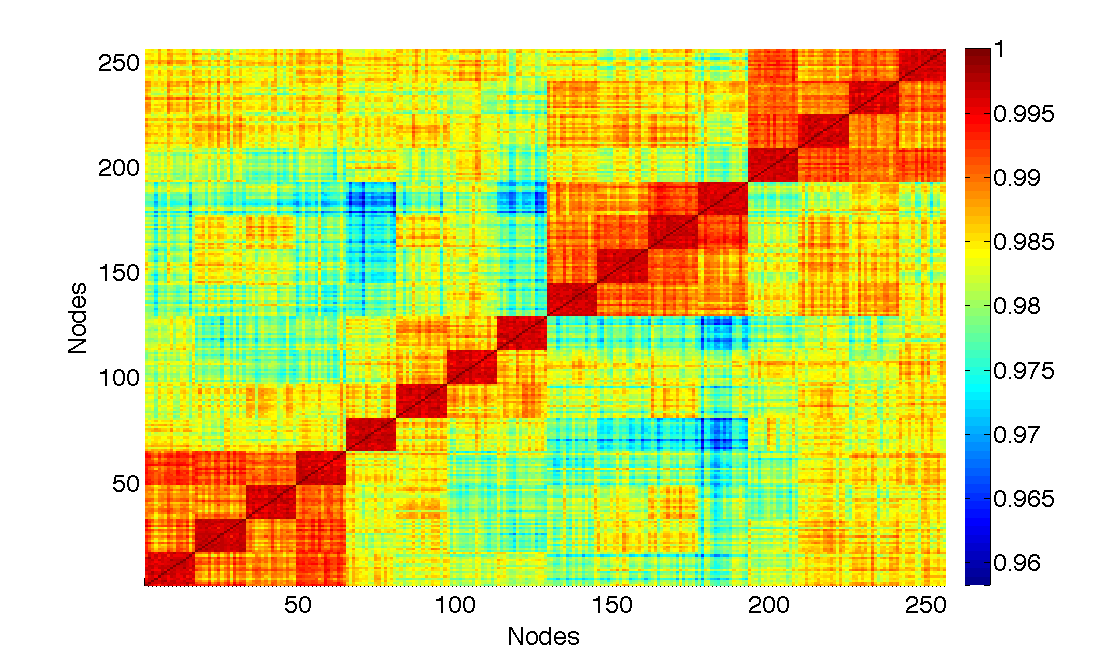} &
\includegraphics[height=0.25\textwidth, width=0.25\textwidth]{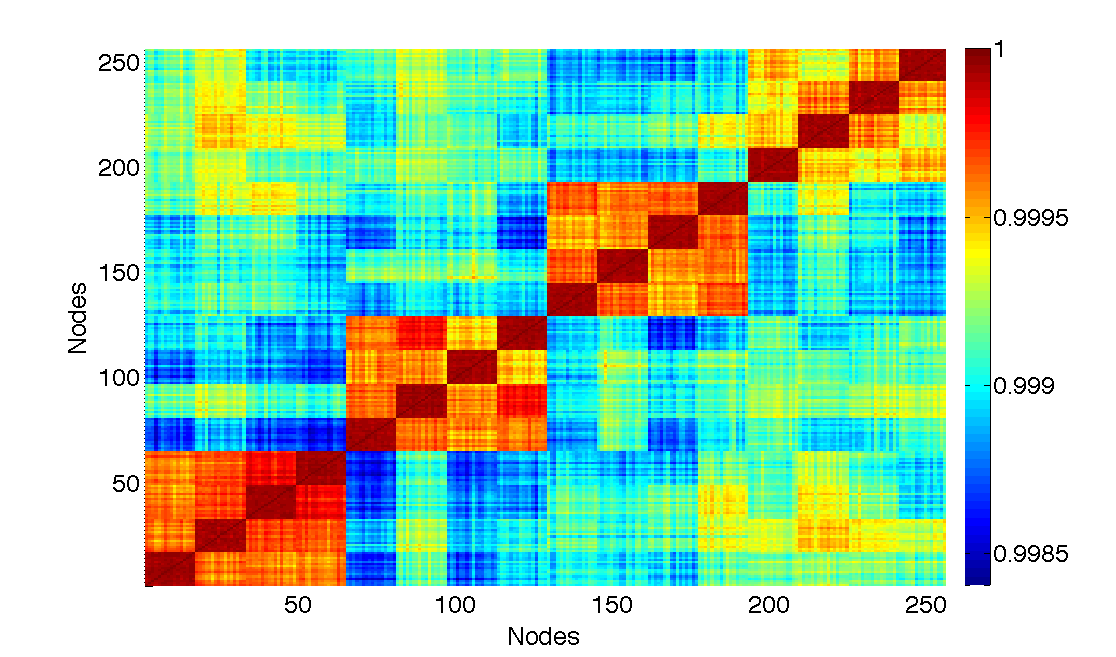}\\
(a)  & (b)   &
(c) &(d)
\end{tabular}
\caption{ (color online) Analysis of the synthetic graph. (a) Hinton diagram of the adjacency matrix. A point is red if an edge exists between nodes at that location; otherwise  it is blue. (b) Eigenvalue spectrum of the two operators. Similarity matrix at $t=1500$ under the (c)  conservative and (d) non-conservative synchronization models. Color indicates how similar two nodes are, with red corresponding to higher and blue to lower similarity.
}
 \label{fig:artificial}
\end{figure*}

While spectral analysis can illuminate some aspects of network structure, simulating dynamics of synchronization offers a more computationally efficient method to identify network structure.  As demonstrated by Arenas et al.~\cite{Arenas06}, nodes in the Kuramoto model synchronize in stages, with smaller units synchronizing before larger units, etc., until the entire network becomes synchronized. These stages reveal the hierarchical community structure of the network.

We quantify the degree of synchronization of nodes $i$ and $j$  at time $t$ using similarity function:
\begin{equation}
s_{ij}(t)=cos\big(\theta_i(t) - \frac{\theta_i^{s}}{\theta_j^{s}}\theta_j({t})\big),
\label{eq:sim}
\end{equation}
\noindent where $\theta_i^{s}$ is the steady state phase of node $i$. The rationale for this metric is that when nodes reach the steady state, further interactions should not change their phases.
In the conservative case, all phases are equal in the steady state; therefore, Eq.~\ref{eq:sim} reduces to  the order parameter $s_{ij}=cos(\theta_i(t) - \theta_j({t}))$ used in \cite{Arenas06}.
%Although the steady state phases in the non-conservative model are not all the same,
As nodes in the same community become more synchronized with each other, their similarity grows. Nodes are maximally similar ($s_{ij}(t)=1$) in the steady state of both synchronization models.

% Next we explore differences between synchronization dynamics
We explore the differences between dynamics of conservative and non-conservative synchronization and the structures that emerge in two example networks.
First, we consider a synthetic network with a fixed hierarchical community structure. While the synthetic network does not have the statistical properties of naturally evolved real-world networks, we study this case to demonstrate that imposing different dynamics on the same graph leads to measurable differences in the structures found by the two synchronization models.
The synthetic network, constructed following the methodology of \cite{Arenas06}, has $N=256$ nodes  evenly divided between four communities, with each community further sub-divided into four equal size sub-communities.
Each node randomly connects to  $z_{1}$ nodes within its sub-community, $z_{1}+z_{2}$  nodes within its community, and $z_{out}$ nodes outside the community. For our experiments, we took $z_{1}=13$, $z_{2}=4$ and  $z_{out}=1$.
Figure~\ref{fig:artificial}(a) shows the hinton diagram of the adjacency matrix of  this network, with red entries indicating the presence and blue the absence of an edge. Dense red blocks correspond to sub-communities at the first level of the hierarchy, and  sparse red blocks to second level communities.
The spectra of $L$ and $R$ operators are shown in Figure~\ref{fig:artificial}(b). Each spectrum contains the eigenvalues of the operator, ranked in descending order, with the largest eigenvalue in the first position. While there are already differences in the spectra of the two operators, these differences become more pronounced in real-world networks characterized by heterogeneous degree distribution.

%KL - how many simulations?
%KL - what is the size of each time step?
We simulate synchronization dynamics in the synthetic network by letting nodes' phases evolve from some initial configuration. We ran 100 simulations of each synchronization model with the initial values of $\theta_i$ drawn from a uniform random distribution $[-\pi,\pi]$ and $\omega_i=0$ $\forall i$, and  $\alpha=\lambda_{max}$ in the non-conservative model.
Figure~\ref{fig:artificial}(c) and (d) show the \emph{similarity matrix} of the network after $t=1500$ iterations under the two models. The matrix represents similarity, $s_{ij}$, of pairs of nodes, with color red corresponding to higher similarity values and blue to lower.  The minimum similarity between any two nodes in the non-conservative system is  $0.998$, compared to $0.958$ for the conservative system. The hierarchical community structure is visible in both similarity matrices.

To find hierarchical community structure of the synthetic network, we execute a hierarchical agglomerative clustering algorithm on the similarity matrix at some time (here, after 1500 iterations). This procedure produces a dendrogram, which can be partitioned into four or 16 clusters. We use normalized mutual information $MI$ to measure how well these clusters reproduce the actual communities~\cite{Danon05}. When $MI=1$, discovered clusters are the actual communities; while for $MI=0$, they are independent of the actual communities.
When we split each dendrogram into four clusters, we find $MI=1.00$ for the conservative model, and $MI=0.83$ for the non-conservative model, while splitting it into 16 clusters, $MI=0.66$ (conservative) and $MI=0.96$ (non-conservative). Non-conservative model appears to identify smaller structures faster and more accurately than the conservative model.

\begin{figure*}[htp]
\begin{tabular}{@{}c@{}c@{}c@{}c@{}}
\includegraphics[height=0.24\textwidth,width=0.24\textwidth]{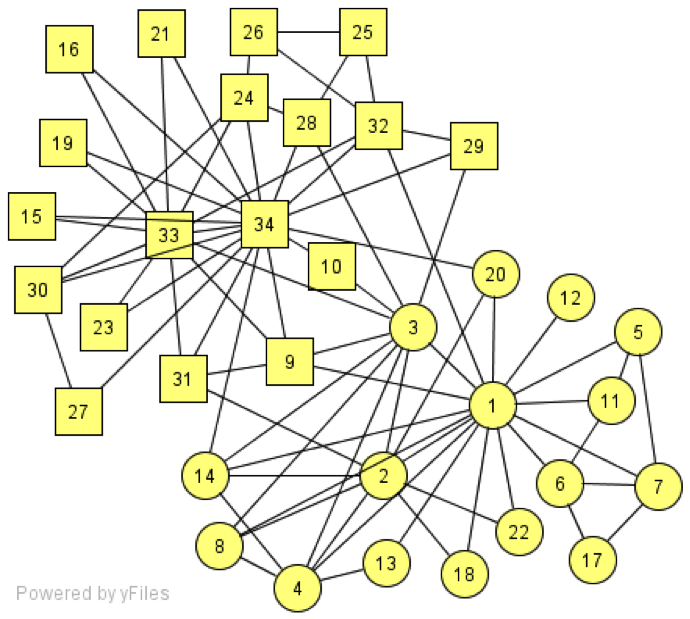}&
\includegraphics[height=0.26\textwidth,width=0.26\textwidth] {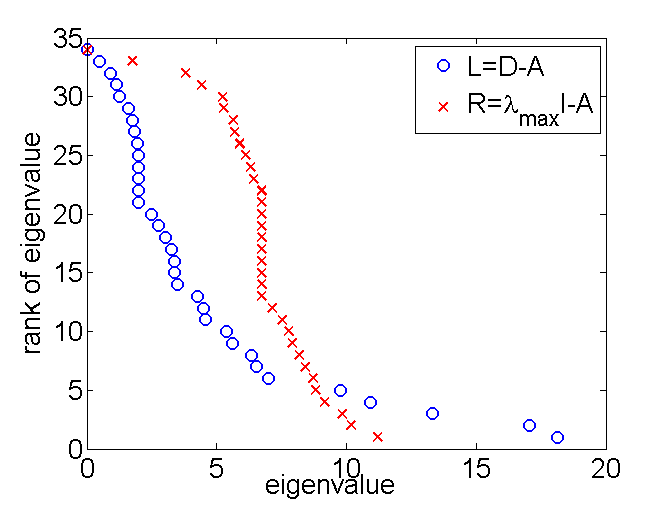} &
\includegraphics[height=0.26\textwidth,width=0.26\textwidth]{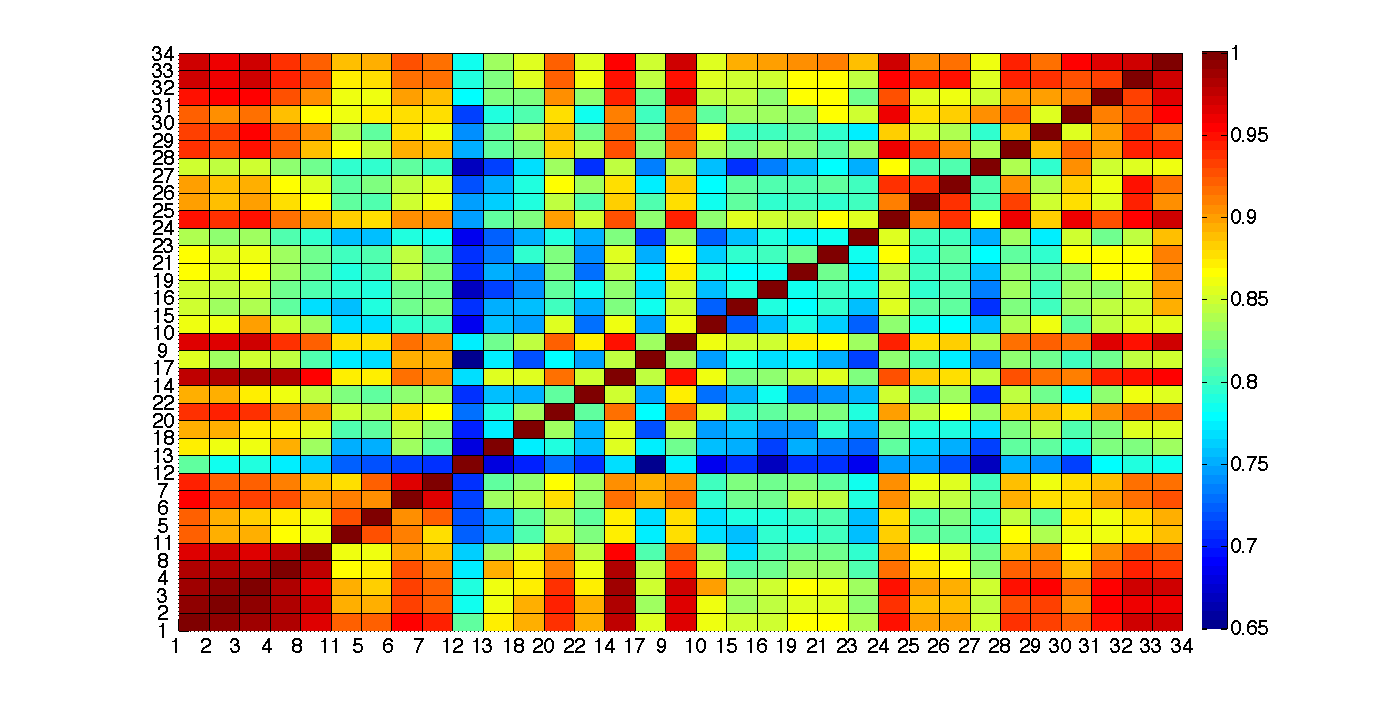} &
\includegraphics[height=0.26\textwidth,width=0.26\textwidth]{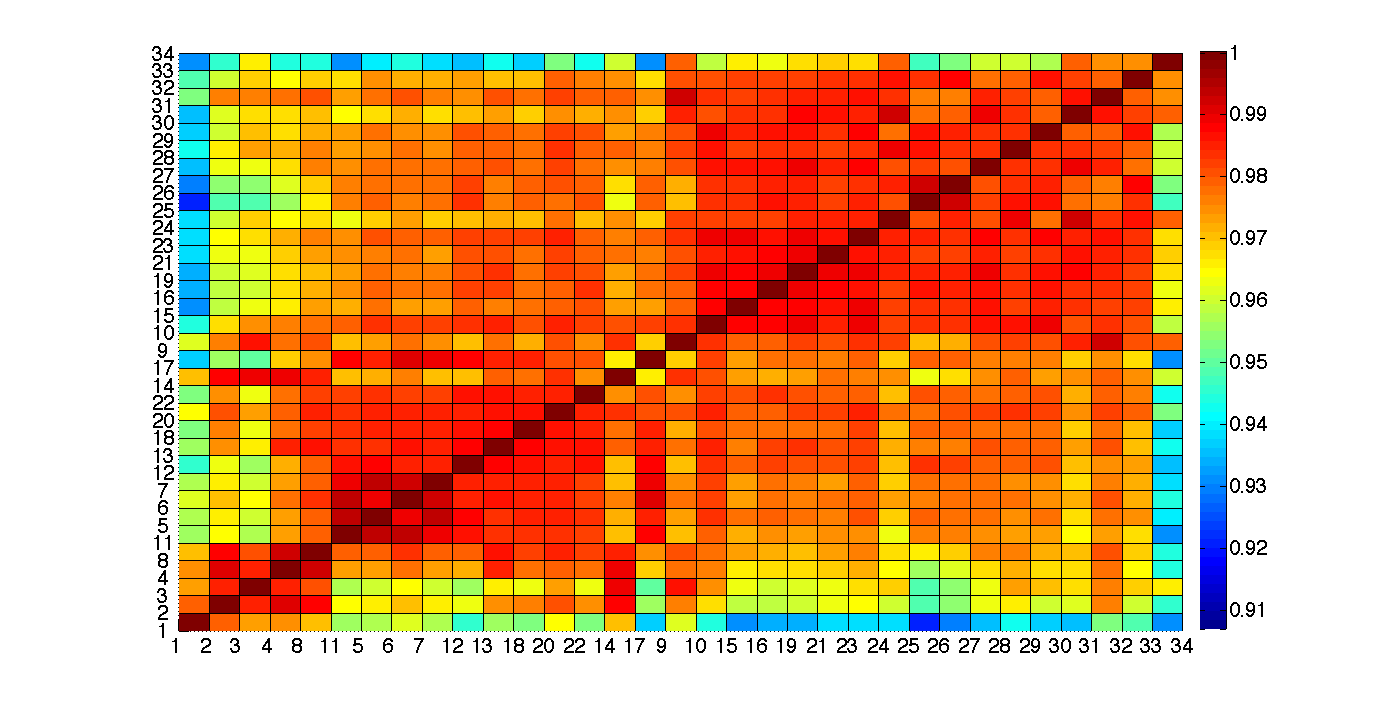} %\hspace{-0.2 in}
\\
(a)  & (b)   & (c) &(d)
\end{tabular}
\caption{(color online) Analysis of the karate club network. (a) Friendship graph. (b) Comparison of eigenvalues of the Laplacian and Replicator operators. Similarity matrix after 1000 iterations of (c) the conservative and (d) non-conservative synchronization models. Color indicates similarity of node pairs, with red corresponding to higher and blue to lower similarity.
}
 \label{fig:karate}
\end{figure*}

Next we study the real-world friendship network of Zachary's karate club~\cite{Zachary}, shown in Fig.~\ref{fig:karate}(a), a widely studied social network benchmark. During the course of the study, a disagreement between the club's administrator and instructor resulted in the division of the club into two factions, represented by circles and  squares, which we take as the actual communities for this network.
There are greater differences between the spectra of $L$ and $R$, shown in Fig.~\ref{fig:karate}(b), than for the synthetic graph with a more homogeneous degree distribution.
The smallest positive eigenvalue of $R$ is larger than that of $L$, implying that the non-conservative model reaches steady state faster than the conservative model.
Figure~\ref{fig:karate}(c) and (d) shows {similarity matrices} after $t=1000$ iterations of the two synchronization models. Minimum similarity in conservative (non-conservative) model is 0.65 (0.91).
%Clearly, nodes are more synchronized in the non-conservative model.

\remove{
\begin{table}
\caption{Normalized mutual information scores of the division of the karate club network into two groups under different synchronization models.}
\label{tbl:karate_nmi}
\begin{tabular}{|l|c|c|c|c|}
  \hline
  \emph{iterations} & \emph{10} & \emph{1000}  & \emph{3000}  & \emph{3899} \\\hline
  \emph{conservative} & 0.046 & 0.046 & 0.046 & 0.733 \\
  \emph{non-conservative} & 0.046 & 0.301 & 1.000 & 1.000 \\
  \hline
\end{tabular}
\end{table}
}

\begin{figure}[htp]
\centering
  \includegraphics[width=0.6\linewidth]{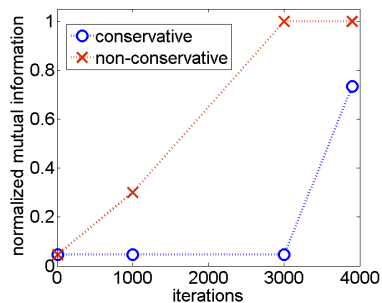}
  \caption{(color online) Normalized mutual information scores of the division of the karate club network into two groups by two synchronization models.}\label{fig:karate_nmi}
\end{figure}

We use a hierarchical agglomerative clustering algorithm to create dengdrograms of the network from the similarity matrices at different times. Both models reveal rich hierarchical structure within the dendrogram, though the two dendrograms are very different. In the conservative model, high degree nodes (hubs) are deeper within the hierarchy, meaning they are more synchronized: 33 and 34,
%21, 9, and 24
in one community and 1, 3 and 2 in the other community. %1 and 3, 2 and 8, 4, 14 in the other community.
Peripheral nodes (such as 13, 18, 22) synchronize later, although nodes 15 and 10 never synchronize with their actual community and are mis-assigned. In the non-conservative model, peripheral nodes %15 and 19, 32 and 29
synchronize first, while the hubs
%1, 3, 33, 34
synchronize later.  Bridging nodes connected to both communities synchronize earlier in the conservative model than the non-conservative model and remain more synchronized.
Figure~\ref{fig:karate_nmi} reports $MI$ scores of communities discovered by the two synchronization models at different times. The non-conservative model identifies communities faster than the conservative model, and the discovered communities are purer.
Under both models community membership of nodes does not change after 3899 iterations.
Similarity continues to increase until every node equally similar to every other node.

In summary, our view of network's community structure depends not only on how its nodes are connected but also on how they interact.
%Different interactions can lead to distinct, possibly incommensurate, views of community structure.
We have explored this issue using models of synchronization in a network of coupled oscillators.
Different interactions lead to distinct operators that govern dynamics of synchronization, each with its own spectral properties and characteristic topological and temporal scales. In practical terms this suggests that to identify communities in real-world networks, algorithms have to take into account the nature of dynamic processes taking place on them. %In other words, the algorithm used to find communities of pages in the Web graph, whose underlying interactions are conservative (e.g., random surfing), is different from one used to find communities in a Twitter follower graph, whose broadcast-based interactions are non-conservative.
%Community detection remains a rich problem.

\bibliographystyle{apsrev}% Produces the bibliography via BibTeX.
\bibliography{references}

\begin{thebibliography}{14}
\expandafter\ifx\csname natexlab\endcsname\relax\def\natexlab#1{#1}\fi
\expandafter\ifx\csname bibnamefont\endcsname\relax
  \def\bibnamefont#1{#1}\fi
\expandafter\ifx\csname bibfnamefont\endcsname\relax
  \def\bibfnamefont#1{#1}\fi
\expandafter\ifx\csname citenamefont\endcsname\relax
  \def\citenamefont#1{#1}\fi
\expandafter\ifx\csname url\endcsname\relax
  \def\url#1{\texttt{#1}}\fi
\expandafter\ifx\csname urlprefix\endcsname\relax\def\urlprefix{URL }\fi
\providecommand{\bibinfo}[2]{#2}
\providecommand{\eprint}[2][]{\url{#2}}

\bibitem[{\citenamefont{Ravasz et~al.}(2002)\citenamefont{Ravasz, Somera,
  Mongru, Oltvai, and Barab\'{a}si}}]{Ravasz2002Hierarchical}
\bibinfo{author}{\bibfnamefont{E.}~\bibnamefont{Ravasz}},
  \bibinfo{author}{\bibfnamefont{A.~L.} \bibnamefont{Somera}},
  \bibinfo{author}{\bibfnamefont{D.~A.} \bibnamefont{Mongru}},
  \bibinfo{author}{\bibfnamefont{Z.~N.} \bibnamefont{Oltvai}},
  \bibnamefont{and} \bibinfo{author}{\bibfnamefont{A.~L.}
  \bibnamefont{Barab\'{a}si}}, \bibinfo{journal}{Science}
  \textbf{\bibinfo{volume}{297}}, \bibinfo{pages}{1551} (\bibinfo{year}{2002}).

\bibitem[{\citenamefont{Rives and Galitski}(2003)}]{Rives03}
\bibinfo{author}{\bibfnamefont{A.~W.} \bibnamefont{Rives}} \bibnamefont{and}
  \bibinfo{author}{\bibfnamefont{T.}~\bibnamefont{Galitski}},
  \bibinfo{journal}{Proc Natl Acad Sci U S A} \textbf{\bibinfo{volume}{100}},
  \bibinfo{pages}{1128} (\bibinfo{year}{2003}).

\bibitem[{\citenamefont{Chung}(1997)}]{Chung1997Spectral}
\bibinfo{author}{\bibfnamefont{F.~R.~K.} \bibnamefont{Chung}},
  \emph{\bibinfo{title}{{Spectral Graph Theory (CBMS Regional Conference Series
  in Mathematics, No. 92)}}} (\bibinfo{publisher}{American Mathematical
  Society}, \bibinfo{year}{1997}).

\bibitem[{\citenamefont{Spielman and
  Teng}(1996)}]{Spielman96spectralpartitioning}
\bibinfo{author}{\bibfnamefont{D.~A.} \bibnamefont{Spielman}} \bibnamefont{and}
  \bibinfo{author}{\bibfnamefont{S.-H.} \bibnamefont{Teng}}, in
  \emph{\bibinfo{booktitle}{In IEEE Symposium on Foundations of Computer
  Science}} (\bibinfo{year}{1996}), pp. \bibinfo{pages}{96--105}.

\bibitem[{\citenamefont{Fortunato}(2010)}]{Fortunato2010Community}
\bibinfo{author}{\bibfnamefont{S.}~\bibnamefont{Fortunato}},
  \bibinfo{journal}{Physics Reports} \textbf{\bibinfo{volume}{486}},
  \bibinfo{pages}{75} (\bibinfo{year}{2010}).

\bibitem[{\citenamefont{Strogatz}(2001)}]{Strogatz01}
\bibinfo{author}{\bibfnamefont{S.~H.} \bibnamefont{Strogatz}},
  \bibinfo{journal}{Nature} \textbf{\bibinfo{volume}{410}},
  \bibinfo{pages}{268} (\bibinfo{year}{2001}).

\bibitem[{\citenamefont{Boccaletti et~al.}(2006)\citenamefont{Boccaletti,
  Latora, Moreno, Chavez, and Hwang}}]{Boccaletti06}
\bibinfo{author}{\bibfnamefont{S.}~\bibnamefont{Boccaletti}},
  \bibinfo{author}{\bibfnamefont{V.}~\bibnamefont{Latora}},
  \bibinfo{author}{\bibfnamefont{Y.}~\bibnamefont{Moreno}},
  \bibinfo{author}{\bibfnamefont{M.}~\bibnamefont{Chavez}}, \bibnamefont{and}
  \bibinfo{author}{\bibfnamefont{D.}~\bibnamefont{Hwang}},
  \bibinfo{journal}{Physics Reports} \textbf{\bibinfo{volume}{424}},
  \bibinfo{pages}{175} (\bibinfo{year}{2006}).

\bibitem[{\citenamefont{Kuramoto}(2003)}]{Kuramoto}
\bibinfo{author}{\bibfnamefont{Y.}~\bibnamefont{Kuramoto}},
  \emph{\bibinfo{title}{Chemical Oscillations, Waves, and Turbulence}}
  (\bibinfo{publisher}{Dover}, \bibinfo{address}{Mineola, NY},
  \bibinfo{year}{2003}).

\bibitem[{\citenamefont{Nishikawa et~al.}(2003)\citenamefont{Nishikawa, Motter,
  Lai, and Hoppensteadt}}]{Nishikawa03}
\bibinfo{author}{\bibfnamefont{T.}~\bibnamefont{Nishikawa}},
  \bibinfo{author}{\bibfnamefont{A.~E.} \bibnamefont{Motter}},
  \bibinfo{author}{\bibfnamefont{Y.-C.} \bibnamefont{Lai}}, \bibnamefont{and}
  \bibinfo{author}{\bibfnamefont{F.~C.} \bibnamefont{Hoppensteadt}},
  \bibinfo{journal}{Phys. Rev. Lett.} \textbf{\bibinfo{volume}{91}},
  \bibinfo{pages}{014101} (\bibinfo{year}{2003}).

\bibitem[{\citenamefont{Arenas et~al.}(2006)\citenamefont{Arenas, Guilera, and
  P\'{e}rez~Vicente}}]{Arenas06}
\bibinfo{author}{\bibfnamefont{A.}~\bibnamefont{Arenas}},
  \bibinfo{author}{\bibfnamefont{A.~D.} \bibnamefont{Guilera}},
  \bibnamefont{and} \bibinfo{author}{\bibfnamefont{C.~J.}
  \bibnamefont{P\'{e}rez~Vicente}}, \bibinfo{journal}{Physical Review Letters}
  \textbf{\bibinfo{volume}{96}}, \bibinfo{pages}{114102+}
  (\bibinfo{year}{2006}).

\bibitem[{\citenamefont{Arenas et~al.}(2008)\citenamefont{Arenas,
  D\'{\i}az-Guilera, Kurths, Moreno, and Zhou}}]{Arenas2008Synchronization}
\bibinfo{author}{\bibfnamefont{A.}~\bibnamefont{Arenas}},
  \bibinfo{author}{\bibfnamefont{A.}~\bibnamefont{D\'{\i}az-Guilera}},
  \bibinfo{author}{\bibfnamefont{J.}~\bibnamefont{Kurths}},
  \bibinfo{author}{\bibfnamefont{Y.}~\bibnamefont{Moreno}}, \bibnamefont{and}
  \bibinfo{author}{\bibfnamefont{C.}~\bibnamefont{Zhou}},
  \bibinfo{journal}{Physics Reports} \textbf{\bibinfo{volume}{469}},
  \bibinfo{pages}{93} (\bibinfo{year}{2008}).

\bibitem[{\citenamefont{Hu et~al.}(2008)\citenamefont{Hu, Li, Zhang, Fan, and
  Di}}]{Hu}
\bibinfo{author}{\bibfnamefont{Y.}~\bibnamefont{Hu}},
  \bibinfo{author}{\bibfnamefont{M.}~\bibnamefont{Li}},
  \bibinfo{author}{\bibfnamefont{P.}~\bibnamefont{Zhang}},
  \bibinfo{author}{\bibfnamefont{Y.}~\bibnamefont{Fan}}, \bibnamefont{and}
  \bibinfo{author}{\bibfnamefont{Z.}~\bibnamefont{Di}}, \bibinfo{journal}{Phys.
  Rev. E} \textbf{\bibinfo{volume}{78}}, \bibinfo{pages}{016115}
  (\bibinfo{year}{2008}).

\bibitem[{\citenamefont{Danon et~al.}(2005)\citenamefont{Danon, Díaz-Guilera,
  Duch, and Arenas}}]{Danon05}
\bibinfo{author}{\bibfnamefont{L.}~\bibnamefont{Danon}},
  \bibinfo{author}{\bibfnamefont{A.}~\bibnamefont{Díaz-Guilera}},
  \bibinfo{author}{\bibfnamefont{J.}~\bibnamefont{Duch}}, \bibnamefont{and}
  \bibinfo{author}{\bibfnamefont{A.}~\bibnamefont{Arenas}},
  \bibinfo{journal}{J. Statistical Mechanics: Theory and Experiment}
  \textbf{\bibinfo{volume}{2005}}, \bibinfo{pages}{P09008}
  (\bibinfo{year}{2005}).

\bibitem[{\citenamefont{Zachary}(1977)}]{Zachary}
\bibinfo{author}{\bibfnamefont{W.~W.} \bibnamefont{Zachary}},
  \bibinfo{journal}{Journal of Anthropological Research}
  \textbf{\bibinfo{volume}{33}}, \bibinfo{pages}{452} (\bibinfo{year}{1977}).

\end{thebibliography}

\end{document}